\documentclass[twocolumn,showpacs,preprintnumbers,amsmath,amssymb]{revtex4}
\usepackage{tabularx,graphicx}

\usepackage{color}
\usepackage{hyperref}
\hypersetup{
    colorlinks=true,
    linkcolor=blue,
    filecolor=blue,      
    urlcolor=blue,
}

\usepackage{color}

\usepackage{ulem}   % to strike things out
\begin{document}
%\documentstyle[aps]{revtex}
%\documentstyle[preprint,aps]{revtex}
%\begin{document}

\newcommand{\beq}{\begin{equation}}
\newcommand{\eeq}{\end{equation}}
\newcommand{\beqn}{\begin{eqnarray}}
\newcommand{\eeqn}{\end{eqnarray}}
\newcommand{\bmath}{\begin{subequations}}
\newcommand{\emath}{\end{subequations}}
\newcommand{\bra}[1]{\langle #1|}
\newcommand{\ket}[1]{|#1\rangle}

%\draft
\title{Flux trapping in   superconducting hydrides under high pressure}

\author{J. E. Hirsch$^{a}$  and F. Marsiglio$^{b}$ }
\address{$^{a}$Department of Physics, University of California, San Diego,
La Jolla, CA 92093-0319\\
$^{b}$Deparrtment of Physics, University of Alberta, Edmonton,
Alberta, Canada T6G 2E1}

\begin{abstract} 
High temperature conventional superconductivity in hydrogen-rich materials under high pressure has been reportedly found in twelve different compounds in
recent years. However, the experimental evidence on which these claims are based has recently been called into question.
Here we discuss the measurement of  trapped magnetic flux, that should establish definitively that these materials are indeed high temperature  superconductors. Its absence would  confirm   claims to the contrary.
\end{abstract}
\pacs{}
\maketitle

\section{introduction}
The paper ``Conventional superconductivity at 203 kelvin at high pressures in the sulfur hydride system,'' published in 2015 \cite{sh3}, launched a revolution
in the field of superconductivity. Following in its footsteps \cite{sh3,sh32}, eleven other hydrogen-rich compounds under high pressure have been found
to be high temperature conventional superconductors based on experimental
evidence:  $PH_x$ above 100~K \cite{eremetsp}, 
$LaH_x$ at 250~K \cite{eremetslah,zhaolah}, above 260~K \cite{hemleylah} and above
550~K \cite{hemleylah2}, $YH_x$   at 243~K \cite{yttrium2,yttrium,yttriumdias}, $ThH_x$ at 161~K \cite{thorium}, $PrH_x$ at 9~K \cite{pr}, 
$LaYH_x$ at 253~K \cite{layh10}, $CSH$ at room 
 temperature \cite{roomt}, $CeH_x$ above 120~K \cite{ceh}, $SnH_x$ at 70~K \cite{snh}, $BaH_x$ around 20~K \cite{bah}, and $CaH_x$ around 215~K \cite{cah,cah2}.
Many more such materials have   been determined to be conventional 
high temperature superconductors based on theoretical evidence \cite{semenok}. As a consequence, a tectonic shift has  taken hold in the field of superconductivity
during the last five years   \cite{review1,review2,theory1,theory2,theory3,struzhkin,nsf}: (1) no longer are the highest temperature superconductors expected to be found among so-called ``unconventional superconductors'' \cite{norman}, as had been
the case at least since the discovery of the cuprates in 1986 \cite{cuprates},
and (2) no longer is the conventional theory of superconductivity   regarded to be unhelpful in the search for new
superconducting materials \cite{matthias}. Quite the contrary, materials theory of the high pressure hydrides has been the driving force 
and guiding light in the long and finally successful quest to 
  find superconductivity close to and even at room temperature \cite{roomt}.
  
%Despite these groundbreaking developments there are however  the facts  that 
Despite this overwhelming evidence, remaining gadflies in the ointment are that
 (a) some of the experimental evidence that  shows that these materials are high temperature
superconductors has recently been called into question \cite{hm,eu,hm2,dc,hm3,hm4,flores}, and (b) the internal consistency of the conventional theory of superconductivity  as well as its ability to
explain the most fundamental property of superconductors, the Meissner effect, have recently been called into question \cite{joule,meissner}.
In this paper we propose that it should be straightforward to present  incontrovertible evidence of the reality of 
conventional superconductivity in high pressure hydrides, if indeed it is real, thus dispensing with both of these criticisms in one fell swoop.

\section{the concept}

 No experimental evidence for magnetic flux expulsion, i.e. the Meissner effect, the hallmark of superconductivity,  has been reported so far for any hydride, rather the opposite \cite{expulsion}. But, it is a fact  that there are some 
bona-fide superconductors that do not expel any magnetic flux \cite{nusran18}. Instead, we propose that 
the experimental observation that would definitively prove the existence of superconductivity in these materials is flux trapping. Not all superconductors will trap magnetic flux. However, we show in this paper that if the one 
experiment \cite{nrs} that was claimed to ``{\it unequivocally confirming the existence of
superconductivity}'' \cite{nrs2} in a hydride was indeed valid, it follows that hydrides should trap
very significant magnetic flux, easily detectable experimentally even under the very challenging experimental conditions of high pressure experiments.

When a superconductor traps flux, it exhibits a magnetic moment in the absence of an externally applied magnetic field.
That magnetic moment is essentially time independent. It originates in  electric current flowing in the material that does not decay with time:
therefore, this current must be a supercurrent, that only superconductors can carry. Thus, the measurement of trapped magnetic flux in a material suspected to be a superconductor
establishes that the material is indeed a superconductor. 
For example, as discussed by K. A. M\"uller and coworkers,  in  early experiments on  high $T_c$ cuprates the samples were cooled in a field which was then switched off and a remnant magnetization was observed,
and the authors concluded that   ``{\it This remanent magnetization results from
flux trapping, and it is a proof of superconductivity in its
own right.}'' \cite{muller}

Not all superconductors trap magnetic flux. Those that do are type II superconductors and in addition are
``hard superconductors'', which means there is a strong pinning force that prevents magnetic vortices from moving, which would dissipate energy and lead
to supercurrent and magnetic field decay.
Many conventional superconductors are hard superconductors, as are many unconventional superconductors such as cuprates and
pnictides. They can trap magnetic fields that are {\it much}  larger than the lower critical field $H_{c1}$ which is typically under $100$~Oe.
For example, $YBCO$ can trap magnetic fields in excess of $17$~T at 29~K \cite{ybco}, and of more than 2~T at 77~K \cite{ybco2}; $Nb_{0.7}Ti_{0.3}$ traps fields of up to 
$0.4$~T  at 4.2~K \cite{nbti}; $GdBa_2Cu_3O_{7-\delta}$ traps  fields in excess of $3$~T  at 50~K \cite{gd}; $SmBaCuO$ traps fields of more than $8$~T at 40~K  \cite{smbacuo}; $MgB_2$ traps fields of up to $3$~T at $5$~K \cite{mgb2}; 
$Ba_{0.6}K_{0.4}Fe_2As_2$ traps fields of up to $1$~T at 5~K \cite{pnictide}.

A counterpart of flux trapping is that the superconductor will effectively screen large externally applied magnetic fields,
fields that are much larger than the lower critical field $H_{c1}$ at the given temperature.
Both the ability to trap magnetic  flux and to screen external magnetic  fields follow from the existence of a large critical current, which in turn depends on the existence
of strong pinning centers. Therefore,   a superconducting material will be able to trap large magnetic fields {\it if and only if} it can screen large externally applied magnetic  fields.

The ability of sulfur hydride ($H_3S$) under high pressure to screen a large externally applied magnetic field was 
established in a seminal nuclear resonant scattering (NRS) experiment by Troyan et al. \cite{nrs}. An applied
magnetic  field of $0.68$~T was
completely excluded from the interior of a thin superconducting disk for temperatures below $50$~K, and some screening occurred even up to
temperatures of $120$~K. This indicates that (if the experiment was valid)  the material will  trap flux in the same range of temperature, of magnitude similar to the flux that
is excluded when a magnetic field is applied. The resulting magnetic moment of the sample should be easily detectable. 
 
In the following two sections we discuss the experiment in quantitative detail for sulfur hydride, and in the last section we summarize our conclusions. Since all high temperature superconducting hydrides are expected to be governed by the same 
physics, our analysis should apply to any of the 12 high pressure superconducting hydrides so far discovered and any more to come.

\section{the NRS experiment}

%figure 1
        \begin{figure} []
 \resizebox{8.5cm}{!}{\includegraphics[width=6cm]{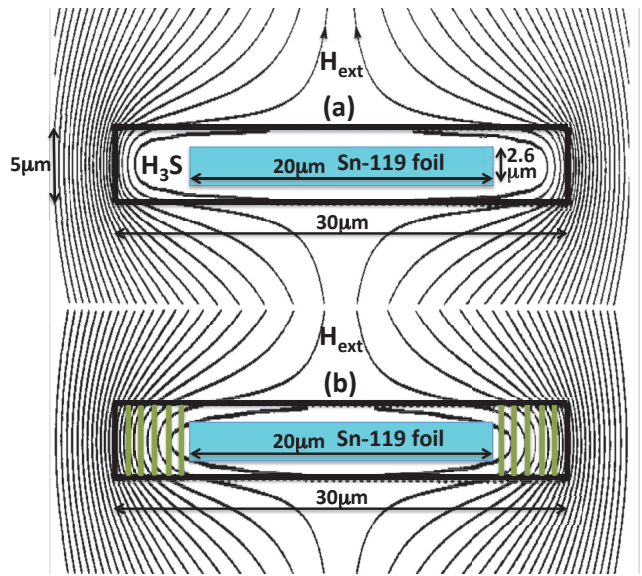}} 
 \caption {Geometry of the NRS experiment with field perpendicular to the sensor film \cite{nrs}.
  The $Sn$ foil is   $20 \ \mu m$ in diameter and $2.6 \ \mu m$ in height, 
 the sulfur hydride superconducting sample is estimated to be $30\ \mu m$ in diameter and $5 \ \mu m$ in  height   \cite{nrs}.
 In (a), we assume the magnetic field only penetrates within the London penetration depth of the lateral surface and the
 superconductor is in the Meissner state. In (b), we assume the field penetrates a much larger distance; this region contains
 pinned Abrikosov vortices (green tubes) and supercurrent flows (see text).
}
 \label{figure1}
 \end{figure} 
 
 Figure~\ref{figure1} shows the geometry of the NRS experiment \cite{nrs}. A magnetic field of magnitude $H_{ext}=0.68$~T was applied
 to an $H_3S$ sample pressurized in a diamond anvil cell (DAC), and the experiment
 established that the magnetic field did not penetrate in the region occupied by a Sn-119 foil
 immersed in the sample. Figure~\ref{figure1} shows
  two scenarios under which this
 can (in principle) be explained.
 
 In the first scenario ((a), top panel), the system is in the Meissner state, and the current that screens the magnetic field in the interior flows in a region within the London
 penetration depth of the surface. In Ref.~\cite{hm3} we showed that this is possible only if $H_3S$ is a `nonstandard superconductor' \cite{hm,hm2,hm3,dc}, with properties
 markedly different from those of standard superconductors; in particular both a {\it much higher} critical current and a {\it much higher} lower critical field
 $H_{c1}$ compared to those
 of standard superconductors are required to explain the data. 
  However, given that theory has convincingly shown  \cite{semenok,review1,review2,theory1,theory2,theory3,nsf} that hydride superconductors are described by conventional BCS-Eliashberg theory that describes standard
 superconductors \cite{tinkham},   the scenario (a) has to be discarded as implausible. That leaves scenario (b) as the only alternative.
 
 In scenario (b),  the lower critical field $H_{c1}$ is much smaller than the applied field of $0.68$~T, as would be the case in any standard
 superconductor where $H_{c1}$ is at most a few hundred Oe. Therefore, the magnetic  field penetrates and the system is in the mixed state in an 
 external annulus of radial thickness up to  approximately $5\mu m$ as shown
 in Fig.~\ref{figure1}(b). There are Abrikosov vortices in the annulus and a supercurrent flows that screens the magnetic field in the interior
 region of diameter $20\ \mu m$ where the $Sn$ foil resides, that thus remains completely  field-free
 in the temperature range $5$~K to $50$~K, as established  by the experiment \cite{nrs}.
 
 The state depicted in Fig.~\ref{figure1}(b) is not a thermodynamic equilibrium state. The supercurrent that flows exerts a radial inward force on the vortices,
 that should cause them to drift inward to the lower energy equilibrium state where there is a uniform distribution of vortices throughout the 
 superconductor \cite{hm3}. The fact that this doesn't happens shows conclusively that there is a large pinning force that prevents the vortices from 
 moving, so that the system remains in this non-equilibrium state at least through the duration of the experiment. 
 The measurements performed in Ref.~\cite{nrs} were measured during a $\sim 25$ min interval for each temperature, and at least six different
 temperatures were measured that showed zero magnetic field in the $Sn$ foil
 in the temperature range $5$~K to $50$~K. This establishes that the non-equilibrium state shown 
 in Fig.~\ref{figure1}(b)) is stable over a period of at least two and a half hours (and probably much longer) in that 
 temperature range.
 
 We will assume in what follows that the Bean critical state model \cite{beanrmp,tinkham,brandt}, widely used to
 describe  hard superconductors, is appropriate to describe the physics of this system. However, even if there were some
 deviations from the Bean model we argue that this would not alter our  conclusions.
 
Within the Bean model, the system shown in Fig.~\ref{figure1}(b) is in the so-called `critical state'. The magnitude of the supercurrent density that flows
(the current density is assumed to be independent of radius) is the
critical current density of this superconductor at the given temperature, which is determined by the strength of the pinning force. The thickness of the region over which supercurrent flows
and vortices exist ($5 \ \mu m$ in Fig.~\ref{figure1}(b)) is determined by the condition that the critical current flowing in that region nullifies the
magnetic field in the interior region where no supercurrent flows.

The experiment shown in Fig.~\ref{figure1} \cite{nrs} was performed under zero field cooling conditions. The system was cooled to a temperature of $5$~K, then the
magnetic field was applied, then the NRS measurements were  performed at increasing temperature. If instead the system is cooled in a magnetic field,
the magnetic field should remain in the interior and vortices will be uniformly distributed in the system.
If the external magnetic field is then removed, the same pinning forces that prevented the flux from penetrating in Fig.~\ref{figure1} will prevent the flux from
leaving the sample: the flux will remain trapped, as shown schematically in Fig.~\ref{figure2}. We discuss this quantitatively in the following section.

\section{trapped flux}

%figure2
         \begin{figure} []
 \resizebox{8.5cm}{!}{\includegraphics[width=6cm]{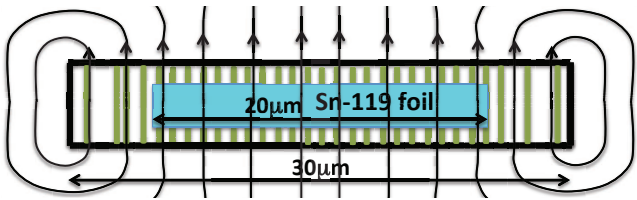}} 
 \caption {Field configuration expected   for   field cooling and then removing the
 applied magnetic field.
}
 \label{figure2}
 \end{figure} 
 
 %figure3
       \begin{figure} []
 \resizebox{8.5cm}{!}{\includegraphics[width=6cm]{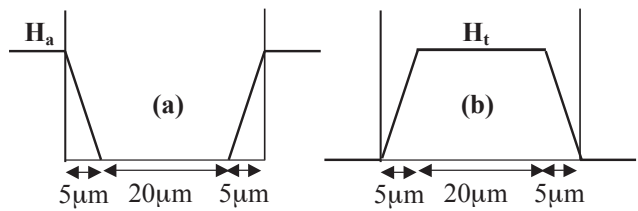}} 
 \caption {(a) Exclusion of external magnetic field $H_a=0.68$~T  from the interior of the sample in the NRS experiment \cite{nrs}, 
 corresponding to the situation shown in Fig.~\ref{figure1}(b). (b) Flux trapping resulting when the system is cooled in the presence of an external field $H_t=0.68$~T and then the external
 field is removed, corresponding to the situation shown in Fig.~\ref{figure2}. The figure assumes no demagnetization, which is exact for a long cylinder.
}
 \label{figure3}
 \end{figure} 
 
When  the system is cooled in a large applied magnetic field the magnetic field should remain in the interior of the sulfur hydride rather than
being expelled. This is always true for type II superconductors cooled in fields larger than the lower critical field $H_{c1}$.
For sulfur hydride, this is true even for applied magnetic fields as small as 20 Oe, as was demonstrated in Ref.~\cite{sh3}, Fig.~4(a).
If the external field is then removed, the field will remain trapped over essentially the entire volume of the superconductor in the presence
of strong pinning centers, as depicted in Fig.~\ref{figure2}.

According to the Bean model, the current flowing is the critical current $J_c$. 
The maximum external field  $H_m$ that can be completely screened out from the center of a long cylinder of radius $r_0$ is given by
\beq
H_m=\frac{4\pi}{c} r_0 J_c
\label{eq1}
\eeq
where $J_c$ is the critical current density.  For the situation of interest here, shown in Fig.~\ref{figure3}(a), the applied magnetic field, $H_{\rm ext}$,
is less than $H_m$, so using Eq.~(\ref{eq1}), and
neglecting demagnetization, the critical current
is given by
\beq
J_c=\frac{c}{4\pi d} H_{ext}
\eeq
with $H_{ext}=0.68$~T the applied magnetic field and $d=5 \ \mu m$  the width of the outer annulus, as depicted in Fig.~\ref{figure3}(a).
This yields for the critical current density
\beq
J_c=1.08\times 10^{7}Amp/cm^2 .
\label{eq3}
\eeq
This estimate agrees with the estimated critical current at low temperatures given   in Ref.~\cite{sh3}.

The same model then predicts that if we field-cool the system in an applied magnetic field of $0.68$~T and
subsequently remove the external field, the applied field will remain trapped over a region of at least
$20\  \mu m$ in diameter, as shown in Fig.~\ref{figure3}(b), with  the same critical current $J_c$ flowing in the outer layer. 
This trapped field should remain over a time period of the same order as the time period over which the system can
keep the applied field out in the zero-field-cooled case, which as discussed above was
established to be greater than 2.5 hours in the NRS experiment.

Let us now estimate the magnetic moment that results from this trapped field.
We can use the measured magnetic moment in sulfur hydride reported in Ref.~\cite{sh3}
for an applied field of $20$ Oe, shown in their Fig. 4(a).
The ZFC curve shows that the system goes from diamagnetic to paramagnetic when the transition 
occurs around $200$~K, and the change in magnetic moment of the sample is approximately
\beq
m\sim 10^{-6} {\rm emu},
\eeq
as given by the difference in ZFC  magnetic moment above and below $T_c$ in Fig. 4(a) of Ref. \cite{sh3}.
The positive signal above $T_c$ seen in Fig. 4(a) of Ref. \cite{sh3} results from a paramagnetic background contribution that
presumably does not change when the sample goes through $T_c$. 

Therefore, these data tell us that a magnetic field of $20$~Oe gives rise to a magnetic moment $m$ in the 
sample that cancels the magnetic field throughout the interior (except within a London penetration depth of the
surface). This implies that if a magnetic field of $0.68$~T is trapped throughout the volume of the sample, it should
generate a magnetic moment approximately $10^{-6}\times (6800/20)\ {\rm emu}=3.4\times 10^{-4} \ {\rm emu}$. 
Finally, we have to correct for the fact that in the outer layer of thickness $5 \ \mu m$ the magnetic field decays
as shown in Fig.~\ref{figure3}(b). This gives then for the estimated
trapped magnetic moment due to the trapped field:
\beq
m_{tr}\sim 3.4\times 10^{-4}\times \left(\frac{12.5}{15}\right)^3 {\rm emu} = 2.0\times 10^{-4} {\rm emu}.
\label{eq5}
\eeq

\begin{figure} []
 \resizebox{8.5cm}{!}{\includegraphics[width=6cm]{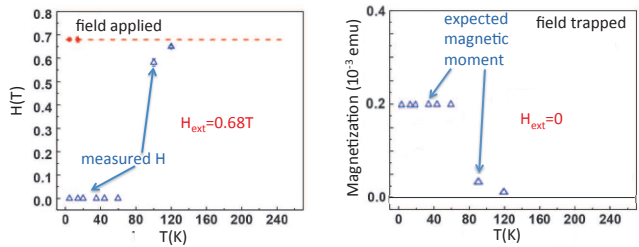}} 
 \caption { Left panel: experimental results for the magnetic field in the interior of $H_3S$ under ZFC with applied field $0.68$~T,
 from Ref.~\cite{nrs}.
 Right panel: resulting magnetic moment under FC to 5 K with applied field $0.68$~T and external field removed, and then increasing 
 the temperature.
 The moment should persist for at least 2.5 hours after the external field is removed.
}
 \label{figure4}
 \end{figure}

The trapped magnetic moment versus temperature is shown in Fig.~\ref{figure4}, right panel. The left panel reproduces the measured magnetic field
in the interior of $H_3S$ under zero field cooling \cite{nrs}.

The sensitivity of the in-situ SQUID magnetometer used in Ref.~\cite{sh3} to obtain the data shown in their Fig. 4(a) was of the order
$10^{-8} \ {\rm emu}$. The expected magnetic moment resulting from trapped flux after application of a 
$0.68$~T magnetic field, Eq.~(\ref{eq5}), is more than
{\it four orders of magnitude} larger. Thus it is  trivial to detect this trapped moment using that
equipment, or  far less sophisticated equipment.

Of course the presence of a background magnetic moment due to the DAC should also be considered. 
However, that contribution will not disappear above $T_c$ as the contribution from the
trapped magnetic moment resulting from supercurrents does. It can also be measured separately by 
performing the experiment without the sample, and subtracted off when the measurement with
the sample is performed.

\section{discussion}
The magnetic moment resulting from trapped flux   Eq.~(\ref{eq5}), shown in Fig.~\ref{figure4}, right panel, should in fact be a lower bound, 
and the real one may well be higher.
This is because the NRS experiment only informed us that the Sn foil remained magnetic-field free, but did not
inform us whether there was also a magnetic field-free region outside of it, which would be the case if the
critical current was larger than Eq.~(\ref{eq3}). If so, the trapped magnetic moment for applied field $0.68$~T 
should be correspondingly higher.
It should also be possible to estimate the maximum value of magnetic moment that can be trapped by
this system for larger applied fields  by   repeating the NRS experiment with larger fields and finding the maximum
external field that keeps the $Sn$ film field-free at a given temperature.

In fact, it would seem that it should have been straightforward to repeat the NRS experiment \cite{nrs} by field-cooling the 
sample in a field of $0.68T$ down to low temperatures, removing the external field, and 
then obtaining the NRS spectra. Observation of quantum beats under those conditions would have
provided compelling proof that the magnetic field had remained trapped in the superconducting sample.
We urge that this test be done when the NRS experiment is repeated to check its reproducibility.

We also point out that the measurements of magnetization versus magnetic field reported in ref. \cite{sh3} 
are  inconsistent with the NRS measurements \cite{nrs}. Fig. 4c of \cite{sh3} shows a paramagnetic response
for fields larger than $0.05T$ at $T=50K$. This is clearly in contradiction with the observations reported in Ref. \cite{nrs} that
a magnetic field of $0.68T$ was excluded from at least $67\%$ of the sample volume (figs. 4A and S6 of Ref. \cite{nrs} at the same $T=50K$).
The caption 
of Fig. 4c in \cite{sh3} reads ``At higher
fields, magnetization increases due to the penetration of magnetic vortexes''. Instead, the field applied (up to 
$0.2T$) should have remained largely excluded according to the results in \cite{nrs}. To resolve this contradiction, one would have to assume that
the $H_3S$ samples used in refs. \cite{sh3} and \cite{nrs} were qualitatively different.

In this paper we have neglected demagnetization effects resulting from the finite aspect ratio 
of the samples. They will not qualitatively affect our conclusions, as can be inferred for example from
the measurements and calculations in Refs.~\cite{nbti,brandt}.

The measurement described here is much simpler and requires much less sensitivity 
than the far more sophisticated measurements performed in
Refs.~\cite{sh3} and \cite{nrs} that claimed to establish that sulfur hydride is superconducting below 203 K.
Detection of a large magnetic moment resulting from trapped flux that persists over time  will confirm that these materials are 
indeed high temperature superconductors, and is an unavoidable consequence  of the physics of these superconductors
revealed by the  NRS experiment \cite{nrs}, namely that they are hard superconductors with strong pinning forces.

Of course, if for some reason the NRS experiment as performed was flawed, as was suggested as a possibility in
Ref.~\cite{hm3}, on one hand it would not provide evidence that $H_3S$ is a 
superconductor, but on the other hand we would also not know for a fact that   {\it if}  these materials are  superconductors, {\it then}
they are hard superconductors, leaving open
the possibility that they could be soft superconductors that do not trap magnetic flux. If so, we still suggest 
it would be revealing to check for the existence of trapped flux in a non-simply connected geometry, by introducing in the
center of the sample a non-superconducting material so that the superconducting material would form a ring around it.
 In that case, quantized flux should certainly be trapped in the non-superconducting material
as long as the magnetic field is smaller than $H_{c1}$.

Detecting trapped magnetic flux in the absence of applied magnetic field that persists over an extended time period will establish  that
these materials are superconductors, by proving  that
supercurrents that do not decay due to resistance  indeed flow in these materials.
$And$, if these materials are indeed superconductors, there can be no doubt that they are $conventional$ superconductors, since 
they owe their very existence  to the conventional theory of superconductivity \cite{ashcroft,ashcroft2}.

On the other hand, failure to detect magnetic moments resulting from trapped flux 
will establish that either (a) these materials are nonstandard superconductors \cite{hm2,dc,hm3}, with properties
qualitatively different from standard superconductors, that additionally don't trap magnetic flux neither in simply connected nor in
multiply connected geometries, as all other known
 superconductors do; or (b), 
far more likely, that these materials  are not superconductors \cite{hm,eu,dc,hm4}.
Either of the two possibilities will call the ability of the conventional theory of superconductivity to describe
real superconductivity of real  materials into question~\cite{validitygeballe}.

\begin{acknowledgments}
FM 
was supported in part by the Natural Sciences and Engineering
Research Council of Canada (NSERC) and by an MIF from the Province of Alberta.
JEH is grateful to S. Shylin for sharing details of the magnetization measurements reported in \cite{sh3} and for extensive discussions.

\end{acknowledgments}
      

\begin{references}
\bibitem{expulsion} 
   Fig. 4a of ref. \cite{sh3} shows that there is no flux expulsion under field cooling in $H_3S$.
\end{references}
 \end{document}